\documentstyle[twoside,fleqn,espcrc2]{article}

\input{epsf}
\title{Two-point function of strangeness-carrying vector-currents in two-loop
Chiral Perturbation Theory
\thanks{Supported by U.S.-DOE contract DE-FG03-96ER40956
and by the Swiss National Science Foundation (SNF).}}

\author{
S. D\"urr
\address{University of Washington, Physics Department, Seattle,
WA 98195-1560, U.S.A.}
\thanks{
Presenter at QCD'99. Address after 1.10.99: Theory Group, Paul Scherrer
Institute, 5232 Villigen, Switzerland}
\thanks{{\tt durr@phys.washington.edu}}
and
J. Kambor
\address{University of Z\"urich, Institute for Theoretical Physics,
8057 Z\"urich, Switzerland}
\thanks{{\tt kambor@physik.unizh.ch}}
}

\begin{document}

\begin{abstract}
We present our calculation of the correlator $\langle T\{V_sV_s^\dag\} \rangle$
between two external vector-currents with the quantum-numbers of a charged kaon.
The renormalized expression to $O(p^6)$ in $SU(3) \times SU(3)$ standard chiral
perturbation theory is finite and scale-independent. The result is used to
determine, via an IMFESR, the phenomenologically relevant finite
$O(p^6)$-counterterm combination $Q_V$ in a way which is not sensitive to
isospin breaking.
\vspace*{-0.1cm}
\end{abstract}

\hyphenation{author another created paper re-commend-ed ex-peri-ment
ex-peri-men-tal counter-term energy}

\maketitle

\def\J#1#2#3#4{{#1} {\bf #2} (#4), #3}
\def\NPB{{\em Nucl. Phys.} {\bf B}}
\def\NPP{\em Nucl. Phys. (Proc. Suppl.)}
\def\PLB{{\em Phys. Lett.} {\bf B}}
\def\PRD{{\em Phys. Rev.} {\bf D}}
\def\PRL{\em Phys. Rev. Lett.}
\def\CMP{\em Commun. Math. Phys.}
\def\AOP{\em Ann. Phys.}
\def\JPA{{\em J. Phys.} {\bf A}}
\def\JPQ{\em J. Physique (France)}
\def\HPA{\em Helv. Phys. Acta}
\def\ZPC{{\em Z. Phys.} {\bf C}}
\def\JHEP{\em J. High Energy Phys.}

\newcommand{\pad}{\partial}
\newcommand{\pas}{\partial\!\!\!/}
\newcommand{\Dsl}{D\!\!\!\!/\,}
\newcommand{\Psl}{P\!\!\!\!/\;\!}
\newcommand{\Nf}{N_{\!f}}
\newcommand{\hqu}{\hbar}
\newcommand{\ovr}{\over} 
\newcommand{\hal}{{1\ovr2}}
\newcommand{\til}{\tilde}
\newcommand{\pri}{^\prime}
\renewcommand{\dag}{^\dagger}
\newcommand{\<}{\langle}
\renewcommand{\>}{\rangle}
\newcommand{\gaf}{\gamma_5}
\newcommand{\lap}{\triangle}
\newcommand{\trc}{{\rm tr}}
\newcommand{\al}{\alpha}
\newcommand{\be}{\beta}
\newcommand{\ga}{\gamma}
\newcommand{\de}{\delta}
\newcommand{\ep}{\epsilon}
\newcommand{\ve}{\varepsilon}
\newcommand{\ze}{\zeta}
\newcommand{\et}{\eta}
\newcommand{\th}{\theta}
\newcommand{\vt}{\vartheta}
\newcommand{\io}{\iota}
\newcommand{\ka}{\kappa}
\newcommand{\la}{\lambda}
\newcommand{\rh}{\rho}
\newcommand{\vr}{\varrho}
\newcommand{\si}{\sigma}
\newcommand{\ta}{\tau}
\newcommand{\ph}{\phi}
\newcommand{\vp}{\varphi}
\newcommand{\ch}{\chi}
\newcommand{\ps}{\psi}
\newcommand{\om}{\omega}
\newcommand{\psb}{\overline{\psi}}
\newcommand{\etb}{\overline{\eta}}
\newcommand{\psd}{\psi^{\dagger}}
\newcommand{\chd}{\chi^{\dagger}}
\newcommand{\etd}{\eta^{\dagger}}
\newcommand{\etp}{\eta^{\prime}}
\newcommand{\rch}{{\rm ch}}
\newcommand{\rsh}{{\rm sh}}
\newcommand{\lab}{\overline{\la}}
\renewcommand{\i}{{\rm i}}
\newcommand{\qal}{q_\al}
\newcommand{\qbe}{q_\be}
\newcommand{\qmu}{q_\mu}
\newcommand{\qnu}{q_\nu}
\newcommand{\gmunu}{g_{\mu\nu}}
\newcommand{\Afin}{A_{{\rm fin}}}
\newcommand{\Bbar}{\overline B}
\newcommand{\Jbar}{\overline J}
\newcommand{\Bmu}{B_\mu}
\newcommand{\Bnu}{B_\nu}
\newcommand{\Bmunu}{B_{\mu\nu}}
\newcommand{\Talmu}{T_{\al\mu}}
\newcommand{\Talnu}{T_{\al\nu}}
\newcommand{\Tmunu}{T_{\mu\nu}}
\newcommand{\Cmu}{C_\mu}
\newcommand{\Cnu}{C_\nu}
\newcommand{\Cmunu}{C_{\mu\nu}}
\newcommand{\Umunu}{U_{\mu\nu}}
\newcommand{\Fsq}{F_0^2}
\newcommand{\psq}{p^2}
\newcommand{\qsq}{q^2}
\newcommand{\msq}{m^2}
\newcommand{\mo}{m_1^2}
\newcommand{\mt}{m_2^2}
\newcommand{\Mo}{M_1^2}
\newcommand{\Mt}{M_2^2}
\newcommand{\Met}{M_\et}
\newcommand{\Mka}{M_K}
\newcommand{\Mpi}{M_\pi}
\newcommand{\Mesq}{M_\et^2}
\newcommand{\Mksq}{M_K^2}
\newcommand{\Mpsq}{M_\pi^2}
\newcommand{\mesq}{M_\et^2}
\newcommand{\mksq}{M_K^2}
\newcommand{\mpsq}{M_\pi^2}
\newcommand{\beq}{\begin{equation}}
\newcommand{\eeq}{\end{equation}}
\newcommand{\bdm}{\begin{displaymath}}
\newcommand{\edm}{\end{displaymath}}
\newcommand{\bea}{\begin{eqnarray}}
\newcommand{\eea}{\end{eqnarray}}


\section{INTRODUCTION}

Since many years QCD is known to have two complementary descriptions which are
both based on quantum field theoretic principles but end up covering entirely
different energy ranges:
Looking at a generic hadronic spectral function as a function of the Mandelstam
variable $s$, Perturbative QCD (PQCD) applies throughout the asymptotic region
down to values of the order of the tau mass squared.
On the other side, Chiral Perturbation Theory (XPT) to one- or two-loop order
gives a satisfactory description near the immediate threshold region.

Much of the challenge there is in QCD derives from the fact that theoretical
access to the wide energy region in between --~actually the region where most
of the experimental information about low-energy hadronic observables is won~--
can be reached only by combining input from either side in a controlled way.
Traditionally this is achieved by the use of QCD sum rules and other dispersive
methods. In particular the elaborate tool of inverse moment finite energy
sum rules (IMFESR) can be used to minimize the numerical impact of the inherent
uncertainties due to the finite accuracy of experimental data and the use of
the operator product expansion (OPE).
The price to pay, however, is that one has to work out the low-energy behaviour
of the correlator underpinning the spectral function one is interested in.

Below we shall present our result for the correlator $\<T\{V_sV_s\dag\}\>$
between two external vector currents with the quantum numbers of a charged Kaon
valid at order $O(p^6)$ in XPT and show how it can be used to determine, via an
IMFESR, the phenomenological combination $Q_V$ of the finite parts of some
$L^{(6)}$-counterterms.


\section{FRAMEWORK}

The appropriate tool to compute the low-energy representation of
$\<T\{V_sV_s\dag\}\>$ is Chiral Perturbation Theory (XPT) with external currents
\cite{XPT}, where the coupling to an external vector current is introduced by
the minimal substitution
\beq
\pad_\mu U \rightarrow \pad_\mu U - \i [V_\mu,U]
\;,
\label{minsubst}
\eeq
with $V_\mu$ having an arbitrary flavour structure, e.g
\beq
V_\mu\equiv V_{s\mu}=\bar q \ga_\mu {\la_4+\i\la_5\ovr 2\sqrt{2}} q
\;.
\label{Vsdef}
\eeq
The non-anomalous counterterm Lagrangian at order $O(p^6)$ has been constructed
in full generality for chiral $SU(N)$ \cite{BCE99} and the divergent part of
the generating functional at the two-loop level has been given in closed form
in \cite{BCEgenfun}.
By now, two-loop calculations (i.e. calculations to $O(p^6)$ in chiral counting)
are tedious but straight-forward.


\section{RESULT FOR $\<T\{V_s^*V_s\}\>$}

Our calculation of $\<T\{V_sV_s^*\}\>$ \cite{DuKa99} closely parra\-lels the
analogous calculation of the correlator be\-tween flavour diagonal currents
considered in~\cite{GoKa95}, except for the non-equal masses in the loops
making it technically more demanding.
We stress that in the light of the application below the complete correlator
is needed, including its finite parts.

Using a method where one-loop subgraphs are renormalized before the second
loop-integration is performed (thereby keeping intermediate expressions
short) and decomposing the correlator
\beq
\Pi_{V,s}=
\i\!\int\!d^4x\;e^{\i q\cdot x}\<0|T\{V_s^\mu(x)\;V_s^\nu(0)\dag\}|0\>
\label{PiDef}
\eeq
according to
\beq
\Pi_{V,s}=
(\qmu\qnu-\qsq\gmunu)\;\Pi^{(1)}_{V,s}(\qsq)+
\qmu\qnu\;\Pi^{(0)}_{V,s}(\qsq)
\label{PiFash}
\eeq
the result was found to read
\bdm
\begin{array}{l}
\vspace{1mm}\!\!
\Pi^{(1)}_{V,s}=
\Big\{-2(L_{10}^{(0)}+2H_1^{(0)})+
\\
\vspace{1mm}\!\!
{3\i\ovr4\qsq}[T_{\rm fin}^{(1)}(\qsq,\mksq,\mpsq)
\!-\!\Afin(\mksq)\!-\!\Afin(\mpsq)]+
\\
\vspace{1mm}\!\!
{3\i\ovr4\qsq}[T_{\rm fin}^{(1)}(\qsq,\mksq,\mesq)
\!-\!\Afin(\mksq)\!-\!\Afin(\mesq)]
\Big\}
\\
\vspace{1mm}\!\!
+\Big\{
-2\i L_5^{(0)}{(\mksq-\mpsq)\ovr\qsq}\;\cdot
\\
\vspace{1mm}\!\!
[3\Afin(\mpsq)-2\Afin(\mksq)-\Afin(\mesq)]+
\\
\vspace{1mm}\!\!
3\i L_9^{(0)}
[T_{\rm fin}^{(1)}(\qsq,\mksq,\mpsq)+T_{\rm fin}^{(1)}(\qsq,\mksq,\mesq)]+
\\
\vspace{1mm}\!\!
3\i L_{10}^{(0)}
[\Afin(\mpsq)+2\Afin(\mksq)+\Afin(\mesq)]-
\\
\vspace{1mm}\!\!
{3\ovr32\qsq}
[-5\Afin(\mpsq)^2+4\Afin(\mpsq)\Afin(\mksq)+
\\
\vspace{1mm}\!\!
6\Afin(\mpsq)\Afin(\Mesq)+4\Afin(\mksq)^2-
\\
\vspace{1mm}\!\!
12\Afin(\mksq)\Afin(\mesq)+3\Afin(\mesq)^2]-
\\
\vspace{1mm}\!\!
{9\ovr32\qsq}
[T_{\rm fin}^{(1)}(\qsq,\mksq,\mpsq)+T_{\rm fin}^{(1)}(\qsq,\mksq,\mesq)-
\\
\vspace{1mm}\!\!
\Afin(\mpsq)-2\Afin(\mksq)-\Afin(\mesq)]^2-
\\
\!\!
{(\mksq-\mpsq)^2\ovr\qsq}O_V-P_V\qsq-4\mksq Q_V-
\end{array}
\edm
\vspace*{-0.6cm}
\beq
\begin{array}{l}
\!\!
4(2\mksq+\mpsq)R_V
\Big\}\cdot{1\ovr F_0^2}
\end{array}
\label{PiOne}
\eeq
\bdm
\begin{array}{l}
\vspace{1mm}\!\!
\Pi^{(0)}_{V,s}=
\Big\{-
{3\i\ovr4}{(\mksq-\mpsq)^2\ovr q^4}\Bbar(\qsq,\mksq,\mpsq)-
\\
\vspace{1mm}\!\!
{3\i\ovr4}{(\mksq-\mesq)^2\ovr q^4}\Bbar(\qsq,\mksq,\mesq)
\Big\}
\\
\vspace{1mm}\!\!
+\Big\{
2\i L_5^{(0)}{(\mksq-\mpsq)\ovr\qsq}\;\cdot
\\
\vspace{1mm}\!\!
[3\Afin(\mpsq)-2\Afin(\mksq)-\Afin(\mesq)]+
\end{array}
\edm
\bdm
\begin{array}{l}
\vspace{1mm}\!\!
{3\ovr32\qsq}[-5\Afin(\mpsq)^2+4\Afin(\mpsq)\Afin(\mksq)+\qquad\,
\\
\vspace{1mm}\!\!
6\Afin(\mpsq)\Afin(\Mesq)+4\Afin(\mksq)^2-
\\
\vspace{1mm}\!\!
12\Afin(\mksq)\Afin(\mesq)+3\Afin(\mesq)^2]+
\\
\vspace{1mm}\!\!
\ka_{K\pi}\cdot{\mksq-\mpsq\ovr\qsq}\Bbar(\qsq,\mksq,\mpsq)+
\\
\!\!
\ka_{K\et}\cdot{\mksq-\mesq\ovr\qsq}\Bbar(\qsq,\mksq,\mesq)+
\end{array}
\edm
\vspace*{-0.5cm}
\beq
\begin{array}{l}
\!\!
{(\mksq-\mpsq)^2\ovr\qsq}O_V
\Big\}\cdot{1\ovr F_0^2}
\;,
\end{array}
\label{PiNot}
\eeq
where
\bdm
\begin{array}{l}
\vspace{1mm}\!\!
\ka_{K\pi}=-6\i(\mksq-\mpsq)L_5^{(0)}+
\\
\vspace{1mm}\!\!
{3\ovr16}[5\Afin(\mpsq)-2\Afin(\mksq)-3\Afin(\mesq)]+
\\
\vspace{1mm}\!\!
{3\ovr32}[3{(\mksq-\mpsq)^2\ovr\qsq}+2(\mksq+\mpsq)-5\qsq]\;\cdot
\\
\vspace{1mm}\!\!
{(\mksq-\mpsq)\ovr\qsq}\;\Bbar(\qsq,\mksq,\mpsq)+
\\
\vspace{1mm}\!\!
{3\ovr32}[-9{(\mksq-\mesq)^2\ovr\qsq}-2(\mksq+\mpsq)+3\qsq]\;\cdot
\\
\vspace{4mm}\!\!
{(\mksq-\mesq)\ovr\qsq}\;\Bbar(\qsq,\mksq,\mesq)
\\
\vspace{1mm}\!\!
\ka_{K\et}=-6\i(\mksq-\mesq)L_5^{(0)}-
\\
\vspace{1mm}\!\!
{9\ovr16}[\Afin(\mpsq)-2\Afin(\mksq)+\Afin(\mesq)]+
\\
\vspace{1mm}\!\!
{3\ovr32}[-{(\mksq-\mpsq)^2\ovr\qsq}-2(\mksq+\mpsq)+3\qsq]\;\cdot
\\
\vspace{1mm}\!\!
{(\mksq-\mpsq)\ovr\qsq}\;\Bbar(\qsq,\mksq,\mpsq)+
\\
\vspace{1mm}\!\!
{3\ovr32}[3{(\mksq-\mesq)^2\ovr\qsq}-{2\ovr3}(13\mksq-3\mpsq)+3\qsq]\;\cdot
\\
\!\!
{(\mksq-\mesq)\ovr\qsq}\;\Bbar(\qsq,\mksq,\mesq)
\;.
\end{array}
\edm
$O_V, P_V, Q_V, R_V$ denote phenomenological linear combinations
of finite parts of $L^{(6)}$-counterterms which have to be determined from
experimental data.
The definition of the integral functions $A_{\rm fin}, \Bbar, T_{\rm fin}^{(1)},
T_{\rm fin}^{(0)}$ is found in the appendix of \cite{DuKa99}.
Note that the integral functions and the coefficients
$L_i^{(0)}, O_V, P_V, Q_V, R_V$ are finite but depend on the renormalization
scale $\mu$.
The combinations $\Pi^{(1)}_{V,s}, \Pi^{(0)}_{V,s}$ have been checked to be
independent of $\mu$~\cite{DuKa99}.
Finally, we shall mention that (\ref{PiFash},~\ref{PiOne},~\ref{PiNot}) has
also been calculated in ref.~\cite{AmBiTa}.


\section{LOW ENERGY THEOREM}

We have performed various consistency checks on the result
(\ref{PiOne},~\ref{PiNot}) for the correlator
(\ref{PiDef},~\ref{PiFash}) \cite{DuKa99}.
Besides this a certain low energy theorem --~so far known to hold true at
one-loop level in the chiral expansion~-- has been found to stay valid at
two-loop level:
Decomposing the correlator $\Pi_{V,s}\!=\!\Pi_{V,s}(\qmu,\qnu)$ in the new
fashion
\beq
\Pi_{V,s}\!=\!
(\qmu\qnu-\qsq\gmunu)\Pi^{(1+0)}_{V,s}\!(\qsq)+
\qsq\gmunu\Pi^{(0)}_{V,s}(\qsq)
\label{Sdecnew}
\eeq
and combining our result for $\Pi^{(1+0)}_{V,s}(\qsq)$ with the analogous
isospin and hypercharge component of the vector correlator obtained in
\cite{GoKa95}, we found
\bdm
\Delta\Pi_V(\qsq)\equiv
(\Pi^{(1+0)}_{V,3}+3\Pi^{(1+0)}_{V,8}-4 \Pi^{(1+0)}_{V,s})(\qsq)
\edm
\vspace*{-0.4cm}
\beq
=\Delta\Pi_V^{\rm 1-loop}(\qsq)+\Delta\Pi_V^{\rm 2-loop}(\qsq)
\label{LET}
\eeq
to be a combination which is {\em free of any counterterm\/}.
This is a ``low energy theorem'' (LET), i.e. a parameter-free
relation between physical quantities, now established at $O(p^6)$ in XPT.
This shows that, despite the many coupling constants present at this order
\cite{BCE99}, there are still specific flavour breaking combinations of
observables which can be predicted without ambiguity.


\section{$Q_V$ VIA AN IMFESR}

An interesting application of our result
(\ref{PiFash},~\ref{PiOne},~\ref{PiNot})
has become possible as the ALEPH collaboration has released its analysis of
$\tau$-decays into hadronic final states with strangeness \cite{ALEPHrhos}.
In principle, our application repeats the determination of $Q_V$ given in
ref.~\cite{GoKa96}, avoiding, however, its potential sensitivity on isospin
breaking effects~\cite{MW98}.
Technically, this is achieved by considering the (hypercharge free)
flavour breaking difference
\beq
\Pi(q^2)\equiv \Pi_{V,3}^{(1+0)}(q^2)-\Pi_{V,s}^{(1+0)}(q^2)
\;,
\label{defPi}
\eeq
which was checked not to have a kinematic singularity at $\qsq\!=\!0$
\cite{DuKa99}. The physical content of the low-energy representation of the
combination~(\ref{defPi}) is analyzed by relating it to the asymptotic
behaviour obtained by the OPE \cite{Na}.
In the present case this is most conveniently done by means of an IMFESR
\cite{earlyFESR}, the mathematical basis of which is simple:
Unitarity and analyticity imply that the correlator (\ref{defPi}) satisfies
\beq
\oint_{C}ds\;{\Pi(s)w(s)\ovr s^{n+1}}=0,
\qquad n=0,1,\ldots
\label{Cauchy}
\eeq
where C is the contour shown in fig.~1 and where
the weight function $w(s)$ must be analytic inside

\noindent
this contour -- a condition which is satisfied for
\beq
w(s)\equiv(1-x)^3\cdot(1+x+\hal x^2)
\;,\quad
x\equiv{s\ovr s_0}
\;.
\label{wchrumm}
\eeq
The task is now to add up (for the case $n\!=\!0$) the contributions from
the various segments, choosing $s_0$ equal the $\tau$-mass squared:

\begin{figure}[t]
\vspace*{-0.2cm}
\begin{center}
\epsfxsize=7.5cm
\epsffile{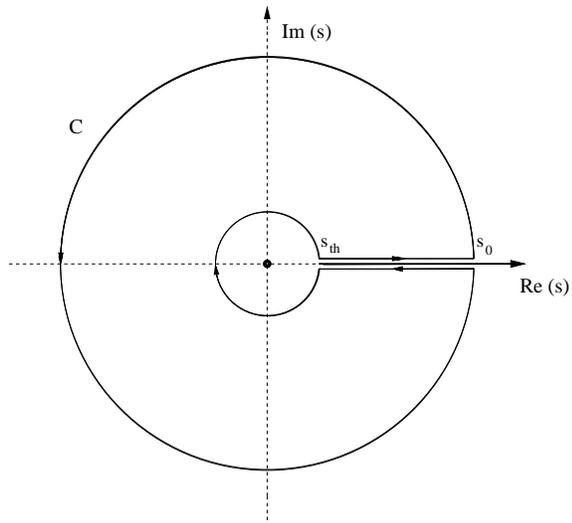}
\vspace*{-0.8cm}
\caption{Integration contour to be used in the IMFESR (\ref{Cauchy}).
The cut on the real axis starts at $s_{\rm th}$ and extends to $\infty$. The
dot at the origin indicates a pole the residue of which is needed to determine
the contribution from the inner circle.}
\end{center}
\label{FigContour}
\vspace*{-1.0cm}
\end{figure}

({\em i\/}) The contribution from the inner circle is $(2\pi/\i)$ times the
residue of the integrand in (\ref{Cauchy}) at $s\!=\!0$, which is $\Pi(0)$,
which we know to $O(p^6)$:
\beq
\Pi(0)=0.0067+{4\ovr F_0^2}(M_K^2-M_\pi^2)Q_V(m_\rho^2)
\;.
\label{PiChPTnum}
\eeq

\begin{table*}[thb]
\begin{center}
\begin{tabular}{lccc}
\hline
$s_{\rm max}$ & $1{\rm GeV}^2$ & $2{\rm GeV}^2$ & $m_\tau^2$\\
\hline
\hline
3-component, $J$=(1)=(1+0) &
$0.0240 \pm 0.0005$ & $0.0254 \pm 0.0005$ & $0.0257 \pm 0.0005$ \\
\hline
$K\pi$-component &
$0.0146 \pm 0.0009$ & $0.0155 \pm 0.0010$ &$0.0155 \pm 0.0010$ \\
$K 2\pi$-component  &
$0.0000 \pm 0.0002$ & $0.0008 \pm 0.0004$ & $0.0011 \pm 0.0006$ \\
$K\eta$- and $K3\pi$-component &
$0.0000 \pm 0.0002$ & $0.0003 \pm 0.0003$ & $0.0004 \pm 0.0004$ \\
\hline
sum s-component, $J$=(1+0) &
$0.0146 \pm 0.0009$ & $0.0166 \pm 0.0011$ & $0.0170 \pm 0.0012$ \\
\hline
3-s, $J$=(1+0) &
$0.0095 \pm 0.0010$ & $0.0088 \pm 0.0012$ & $0.0077 \pm 0.0013$ \\
\hline
\end{tabular}
\end{center}
\caption{Hadronic integrals $B^{(J)}_{V,f}$ for various components $f$ of the
total vector spectral function $\rho^{(J)}_{V}$ [see (\ref{rhodef})].}
\label{tabone}
\end{table*}

({\em ii\/}) The contribution from the straight segments is
$B^{(1+0)}_{V,3-s}\!=\!\int_{s_{\rm th}}^{s_0}ds\;
\rh_{V,3-s}^{(1+0)}(s)w(s)/s$, where
\bdm
\rh_{V,3-s}^{(1+0)}(s)={1\ovr\pi}{\rm Im}\Pi_{V,3}^{(1+0)}(s)\th(s-4\Mpi^2)-
\edm
\vspace*{-6mm}
\beq
{1\ovr \pi} {\rm Im} \Pi_{V,s}^{(1+0)}(s) \th(s-(M_K+M_\pi)^2)
\label{rhodef}
\eeq
is the spectral function of the correlator (\ref{defPi}).
This is the place where the ALEPH data for $\rh_{V,s}(s)$ and $\rh_{V,3}(s)$
get used \cite{ALEPHrhos,ALEPHrho3}.
As one can see from table~1, our weight polynomial (\ref{wchrumm}) proves very
efficient in suppressing the region above $2 {\rm GeV}^2$, where experimental
accuracy deteriorates.

({\em iii}) The contribution from the outer circle is evaluated by means of
the OPE.
Again, our weight polynomial (\ref{wchrumm}) suppressing the potentially
dangerous region $s\simeq s_0$ so strongly proves beneficial:
$D\!=\!2$ operators contribute just $2\pi\i\cdot$ $(8.5\pm3.3)10^{-4}\!$
and $\!D\!\geq\!4$ operators are negligible.

Combining these three inputs, we end up with
\beq
Q_V(\mu=m_\rho)=(1.8 \pm 1.2)\cdot 10^{-5}
\label{QVresult}
\eeq
which is both, relatively precise and model independent.
For details the reader is referred to \cite{DuKa99} and references therein.


\section{SUMMARY AND CONCLUSIONS}

We have presented our result for the strange component of the correlator
between two external vector currents to two loops in Chiral Perturbation
Theory (XPT).
Furthermore, it has been shown how it can be combined with a known expression
to form a new low-energy-theorem, i.e. a parameter-free relation between
chiral correlators with different flavour structure.

The physical application we have sketched is a first-principle-based
determination of the phenomenological combination $Q_V$ of finite parts of the
$L^{(6)}$-lagrangian by means of an inverse moment finite energy sum rule.
This technique allows to combine the low-energy representation of a
correlator (obtained from XPT) and its high energy expansion (from OPE and
PQCD) with experimental data (won in the intermediate region) to pin down
so-far undetermined low energy constants inherent to the chiral representation.
Last but not least, the IMFESR-calcu\-lation makes transparent in which way the
finite parts of the chiral counterterms know, through their values,
about the physics at higher scales.


\end{document}